\documentclass{article}
\usepackage{times}  
\usepackage{lmodern}  
\usepackage{courier}  
\usepackage{graphicx}  
\usepackage{hyperref}  
\usepackage{amsmath, amssymb}  
\usepackage{algorithm, algorithmic}  
\usepackage{float}
\usepackage{enumitem}

\title{Empowering Application Modernization with LLMs: Addressing Core Challenges in Reliability, Security, and Quality}
\author{Ahilan Ayyachamy Nadar Ponnusamy\\ \texttt{ahilanp@gmail.com}}

\begin{document}

\maketitle

\begin{abstract}
AI-assisted code generation tools have revolutionized software development, offering unprecedented efficiency and scalability. However, multiple studies have consistently highlighted challenges such as security vulnerabilities, reliability issues, and inconsistencies in the generated code. Addressing these concerns is crucial to unlocking the full potential of this transformative technology. While advancements in foundational and code specialized language models have made notable progress in mitigating some of these issues, significant gaps remain, particularly in ensuring high quality, trustworthy outputs.

This paper builds upon existing research on leveraging large language models (LLMs) for application modernization. It explores an opinionated approach that emphasizes two core capabilities of LLMs: \textbf{code reasoning and code generation}. The proposed framework integrates these capabilities with human expertise to tackle application modernization challenges effectively. It highlights the indispensable role of human involvement and guidance in ensuring the success of AI-assisted processes.

To demonstrate the framework's utility, this paper presents a detailed case study, walking through its application in a real-world scenario. The analysis includes a step-by-step breakdown, assessing alternative approaches where applicable. This work aims to provide actionable insights and a robust foundation for future research in AI-driven application modernization. The reference implementation created for this paper is available here \href{https://github.com/AhilanPonnusamy/App-Modernization-framework-reference-implementation}{GitHub Repository}.
 
\end{abstract}


\section{INTRODUCTION}

Legacy systems, particularly those developed during the dot-com boom and early phases of digital transformation, represent a critical challenge for modern enterprises. While these systems have been the operational backbone for a couple of decades, they are increasingly hampered by technological obsolescence, scalability issues, and maintenance complexity. Within the scope of this paper, we focus on legacy systems that utilize relatively modern programming paradigms and languages, excluding older systems developed in languages like COBOL and Fortran. The exclusion of older languages is driven primarily by the limited public knowledge and training data available for them, which reduces the effectiveness of large language models (LLMs) in reasoning about or generating code for these technologies.\\

Instead, this paper focuses on modernization efforts for systems developed in more contemporary languages, such as Java and .NET, hereafter referred to as legacy applications. These systems are well suited for leveraging LLMs, given the abundance of available training data and the demonstrated proficiency of LLMs with recent technologies. Additionally, as these applications often operate across the internet, mobile, and edge environments, they face heightened security and compliance risks, making their timely modernization crucial. Code specialized, open source LLMs have matured to a point where they can meaningfully contribute to application modernization without requiring costly and time intensive fine-tuning. This is particularly important for lowering barriers to entry, a key focus of this paper. By leveraging readily available, open source models, organizations can explore modernization with minimal investment in infrastructure or specialized expertise.\\

Another aspect of this paper is the emphasis on accessibility and democratization. The proposed framework is built for enterprise-scale deployment. The reference implementation tool, while optimized to run efficiently on modest hardware like a MacBook Pro, can seamlessly scale for enterprise-level deployments. This makes it feasible for a wide range of users to experiment, customize, and adopt the framework based on their specific needs.\\

The key contributions of this paper are as follows:

\begin{itemize}
    \item \textbf{Addressing LLM Limitations for Legacy Systems:} The paper explores the challenges faced by LLMs in modernizing legacy applications, particularly when direct code translation approaches are employed \cite{pietrini2024bridging}.
    \item \textbf{Leveraging Code Reasoning for Modernization:} By utilizing the improving code reasoning capabilities of LLMs, the framework enables thoughtful and context-aware modernization processes \cite{diggs2024leveraging}.
    \item \textbf{Adopting ``Progressive Prompting'' for Incremental Progress:} An effective technique, ``progressive prompting,'' is introduced to streamline the process of generating, refining, and validating modernized code in incremental steps \cite{wei2024requirements}.
    \item \textbf{Integrating Security Best Practices and IP Protection:} To mitigate risks and ensure intellectual property (IP) protection, the framework incorporates security-focused best practices, making it enterprise-ready \cite{torka2024optimizing}.
    \item \textbf{Democratizing Adoption:} The paper proposes a standardized framework that reduces the complexity of experimentation and adoption. This is particularly relevant for enterprises seeking to explore modernization without incurring significant upfront costs.
    \item \textbf{Promoting Local Deployment:} By leveraging open models, the paper highlights a lightweight, local deployment approach that minimizes the need for specialized hardware and cloud services, lowering the barrier to experimentation and early adoption.
\end{itemize}

This paper builds on existing research to present a pragmatic and accessible approach to AI-assisted application modernization. By focusing on a framework that integrates LLM capabilities with human oversight, it bridges the gap between theoretical potential and practical implementation, offering actionable insights for enterprises seeking to modernize legacy systems.

\section{RESEARCH METHOD}

To address the challenges and complexities outlined above, this paper adopts a structured approach divided into two key subcategories:

\subsection{Requirements Generation for Legacy Applications}

The first step involves leveraging LLMs to analyze legacy application code and generate comprehensive functional requirements. These requirements are then reviewed and refined by SMEs to ensure accuracy, relevance, and alignment with modernization objectives. This step serves as the foundation for building a modernized application that aligns with both current and future business needs.

\subsection{Development of the New Application Based on Refined Requirements}

Using the reviewed and refined requirements, the framework guides the generation of a new application in a step-by-step manner. This process focuses on translating the modernized requirements into functional code, utilizing the reasoning and code generation capabilities of LLMs.\\

The overall process is supported by a robust framework and a purpose-built reference implementation tool designed to standardize and streamline the modernization workflow. This framework provides clear guidance for each phase of the process, ensuring consistency, repeatability, and the integration of human oversight at critical decision points.

By organizing the research into these two subcategories and grounding the process in a practical framework and toolset, this paper provides a comprehensive, accessible, and actionable approach to AI-assisted application modernization.

\section{THE FRAMEWORK AND THE TOOLl}

The framework aims to simplify and streamline the modernization of legacy applications by providing a step-by-step approach, from generating requirements to producing modern, enhanced code for the target platform. It focuses on two key areas:

\begin{itemize}
    \item \textbf{Requirements Generation from Legacy Applications}
    \item \textbf{Development of a New Application Based on Refined Requirements}
\end{itemize}

\subsection{Requirements Generation from Legacy Applications}
The following interaction diagram illustrates the requirements generation from legacy applications process in detail.

\vspace{-10pt}
\begin{figure}[H]
    \centering
    \includegraphics[width=0.75\textwidth]{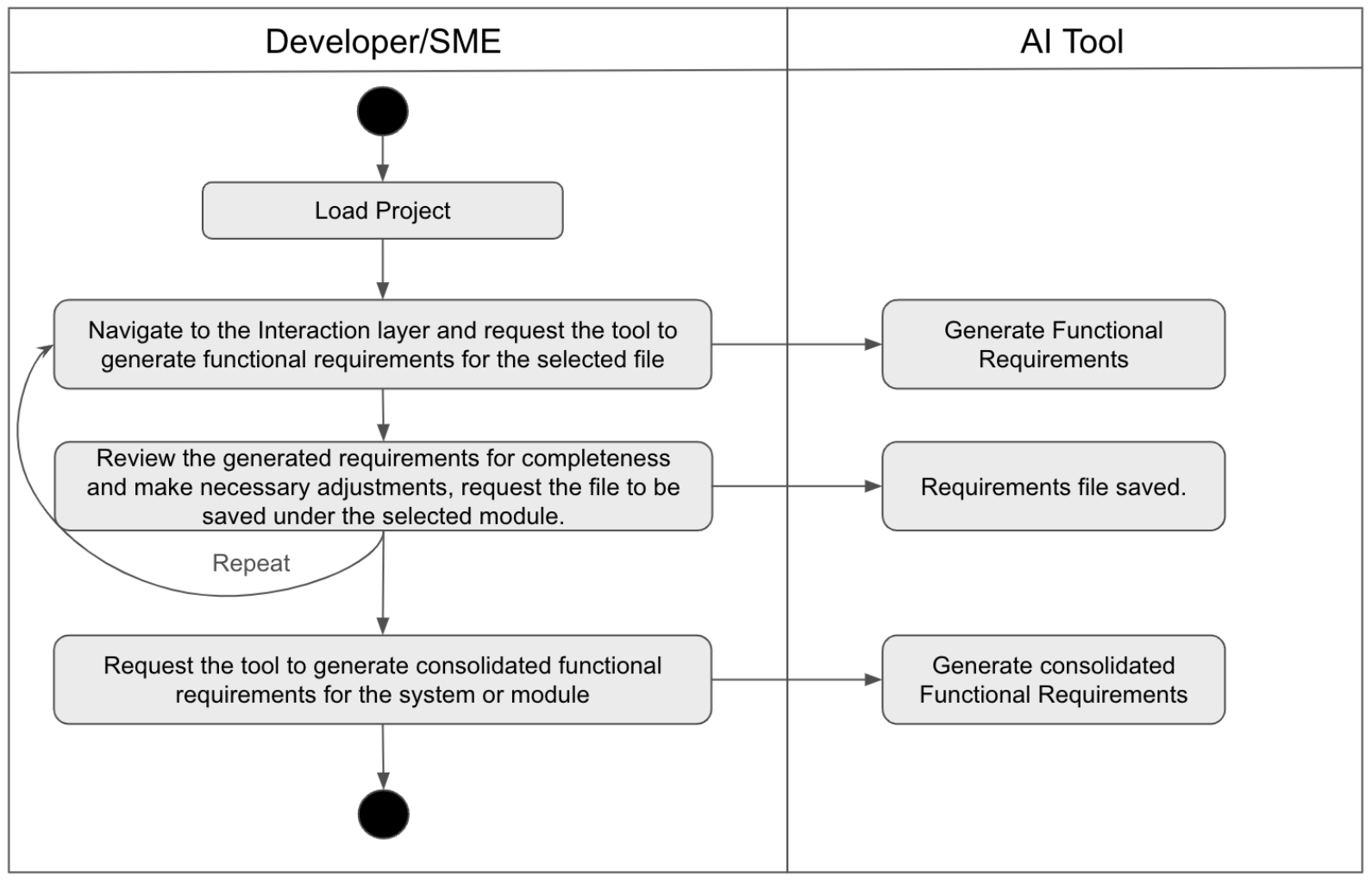}
    \caption{Interaction diagram showing the requirements generation process.}
    \label{fig:req-gen-process}
\end{figure}

The framework recommends starting with the \textbf{Interaction Layer}, which serves as the entry point for users (via UI) and other systems (via APIs or Batch scripts). Since this layer typically captures most of the functional requirements, it forms a solid starting point.\\

Next, attention shifts to the \textbf{Business Logic Layer}, where specific calculations and business rules are implemented. Finally, the process moves to the \textbf{Data Layer} or \textbf{Configuration Files}, which store system level information, properties, and lists. By covering these three layers, the framework ensures that the core functional requirements of the legacy application are captured, providing a strong foundation for modernization.

\vspace{-10pt}
\begin{figure}[H]
    \centering
    \includegraphics[width=0.9\textwidth]{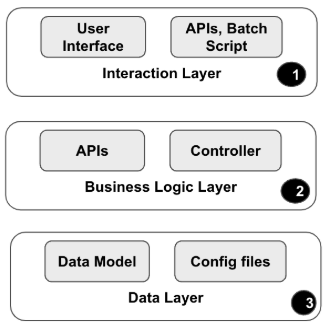}
    \caption{Recommended steps for Requirements Generation.}
    \label{fig:req-gen-steps}
\end{figure}

It's worth nothing that while legacy applications may have existing requirements or design documents, these are often outdated and may not fully represent the current state of the application. Extracting requirements directly from the application code is more reliable, as it reflects the latest system behavior. Comparing these extracted requirements with existing documentation can help identify gaps and discrepancies, which can then be addressed.\\

This approach also provides an opportunity to enhance the application beyond its current state. Organizations can include additional features, address existing bugs, or modernize certain functionalities. Unlike direct code translation methods, this method allows for a broader scope, enabling the creation of a more robust and feature-rich target application.

\subsection{Development of a New Application Based on Refined Requirements}
The following interaction diagram illustrates the new application development process in detail.

\vspace{-10pt}
\begin{figure}[H]
    \centering
    \includegraphics[width=0.9\textwidth]{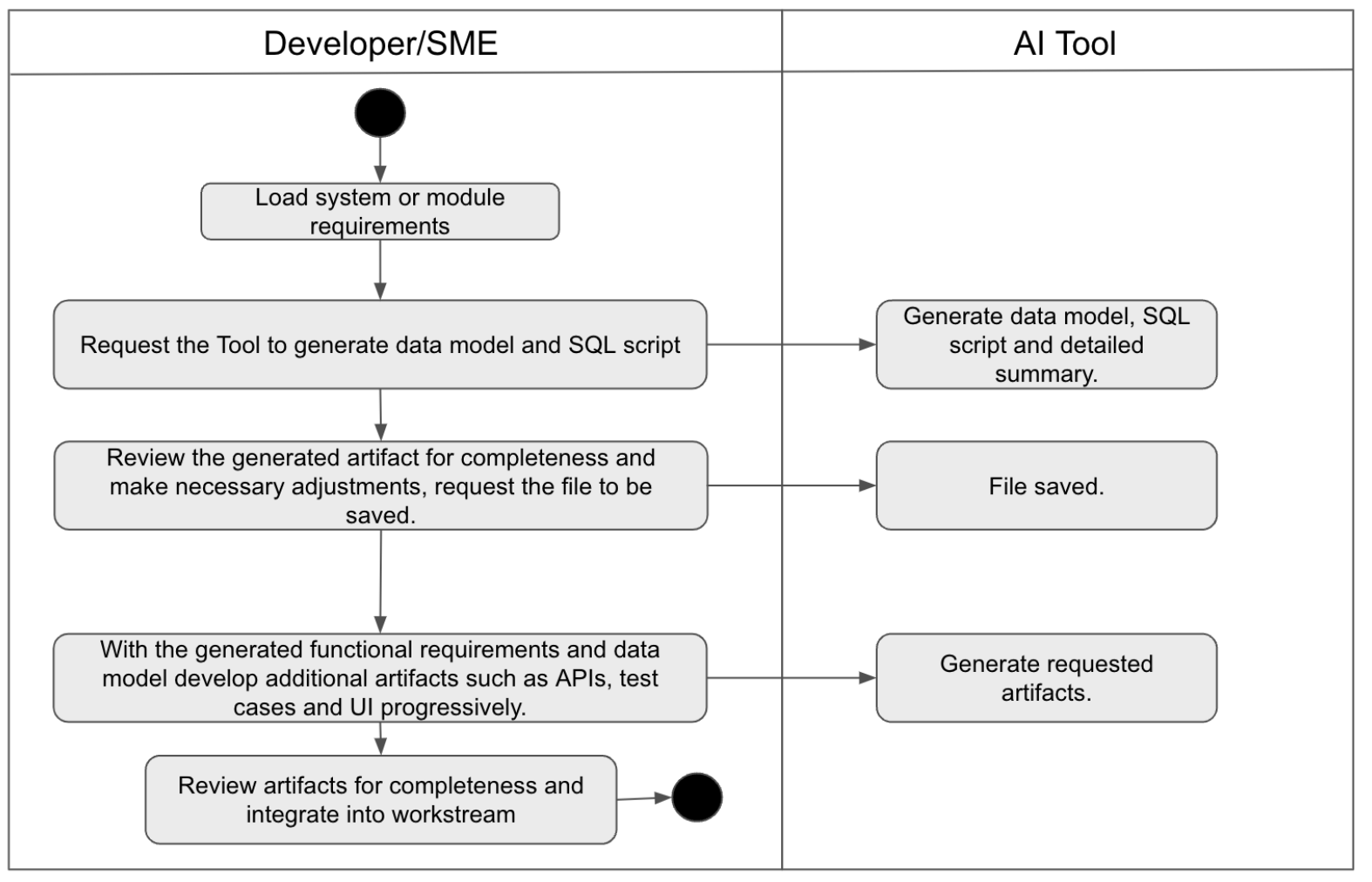}
    \caption{Interaction diagram showing the new application development process.}
    \label{fig:app-dev-process}
\end{figure}

With the requirements generated in the previous phase, the development of the new (modern) application can be approached either module by module or for the entire application as a whole. Similar to the requirements generation phase, the framework emphasizes a step-by-step process, with human review and adjustments incorporated at every stage.\\

Through trial and error, the most effective approach for generating code for the modernized application using LLMs has proven to be a \textbf{layer-by-layer} approach, but in reverse order compared to the previous phase. Specifically:

\begin{itemize}
    \item \textbf{Start with the Data Layer}: Define and implement database schemas, properties, and configurations to establish a foundation for the application.
    \item \textbf{Move to the Business Logic Layer}: Develop the application’s core logic and calculations, ensuring the system meets functional requirements.
    \item \textbf{Conclude with the Interaction Layer}: Design and implement user interfaces and APIs, ensuring seamless interaction for users and external systems.
\end{itemize}

This methodical, reverse layer approach ensures that each layer builds upon a stable and well defined foundation, enabling a smoother and more efficient modernization process.

\vspace{-10pt}
\begin{figure}[H]
    \centering
    \includegraphics[width=0.75\textwidth]{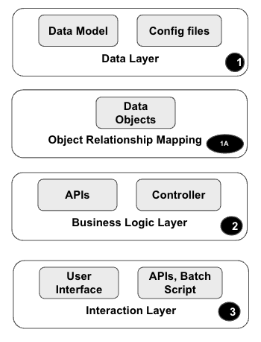}
    \caption{Recommended steps for Application (code) generation.}
    \label{fig:app-gen-steps}
\end{figure}

While the framework discussed earlier can be manually implemented to achieve modernization goals, manual adoption requires significant expertise and discipline. To ensure a standardized, consistent, and efficient process that delivers optimal results, it is recommended to build a dedicated tool. Such a tool would automate and enforce the framework's principles, reducing the room for error and improving outcomes.

\subsection{Key Aspects of the Reference Implementation Tool}
\begin{itemize}
\item \textbf{Specialized Code Model (LLM) of Choice:} The core of the tool is a specialized code model selected for its capability to handle code generation and reasoning tasks effectively. This paper emphasizes lowering the barrier to entry for organizations, favoring open models available under a permissive Apache 2.0 license. After evaluating several options, the \texttt{granite-8b-code-instruct-128k} model was chosen due to its transparency in detailing the composition and source of its training data over alternatives like Code Llama and Codestral models~\cite{ibm2024granite, ibm2024responsible}.  The author is affiliated with Red Hat, a subsidiary of IBM; however, this selection was based solely on model transparency. Granite is a decoder-only model, which, while not the preferred architecture as explained in~\cite{wang2021codet5}, has demonstrated strong performance for our needs. The tool's reference implementation is model-agnostic, allowing organizations to switch to their preferred models.\\

\item \textbf{Web Application for Framework Implementation:} The tool is envisioned as a web application designed to operationalize the framework outlined in this paper. By automating the progressive steps of the framework, the tool ensures consistency and simplifies adoption across teams.\\

Prompt engineering is a crucial element of the framework, and minor variations in prompts can produce vastly different outputs, complicating code consistency and maintenance. The framework relies on \textbf{progressive prompting}~\cite{wei2024requirements}, where each step builds on the previous, necessitating a structured approach to avoid errors and maintain the logical flow.\\

The web application provides a user-friendly interface, enabling users to follow the prescribed steps intuitively. This mitigates the need for deep expertise in LLMs or prompt engineering while fostering broader adoption.
\end{itemize}
It is important to highlight the key difference between this framework and AI code assistant IDE (Integrated Development Environment) plugins. AI code assistant plugins typically offer two main features:

\begin{enumerate}
    \item Chat interface.
    \item Code completion.
\end{enumerate}

The code completion feature is highly beneficial, as it generates real-time code suggestions that enable faster development and occasionally introduces developers to new problem-solving approaches they might not have considered. However, the chat interface introduces variability, as the outcomes depend heavily on how questions (prompts) are phrased by the user.\\

This framework addresses this issue by eliminating user-dependent prompting. Instead, it embeds the prompts directly into the tool, ensuring consistent results. This approach enhances both reliability and maintainability of the development process.\\

During the case study, it was observed that the code generated by the reference implementation tool was significantly augmented by code completion, which helped developers address syntax errors and extend the generated code more effectively. As noted in~\cite{sergeyuk2024ai}, developer opinions on AI assistants vary widely. However, leveraging the reference implementation tool in conjunction with the code completion features of AI assistants appears to enhance overall experience and quality.

\subsection{Implementation of the Tool}
The reference tool implementation was itself generated using an LLM via progressive prompting, with minimal human intervention, demonstrating the framework’s capability in practical use. The following benefits highlight its value:

\begin{itemize}
    \item \textbf{Lower Barrier to Entry}: Utilizes a code specialized open model deployable on local hardware with minimal resource requirements.
    \item \textbf{Enhanced Security and IP Protection}: Supports air-gapped deployments, addressing key concerns around intellectual property (IP) protection and security risks like prompt injection and jailbreak attacks \cite{gupta2023threatgpt}.
    \item \textbf{Consistency Through Prompt Engineering}: Implements standardized prompts to ensure consistency and efficiency across teams, addressing some best practices proposed in \cite{perry2023users}.
\end{itemize}

The tool complements the framework by automating critical steps, ensuring consistent implementation, and addressing challenges associated with manual adoption. As detailed in the case study section, the reference implementation tool provides a simple yet powerful baseline, empowering organizations to modernize their legacy systems effectively and securely leveraging AI.

\subsection{Ensuring Quality and Verification in AI-Assisted Code Generation}
A critical consideration in AI-assisted requirements and code generation is maintaining quality and security throughout the process. Studies indicate that the use of AI assistants can increase security vulnerabilities by 10\% \cite{sandoval2023lost}. To mitigate these challenges, the proposed framework and its reference implementation adopt a \textbf{human-centered protection approach}. \\
\begin{itemize}
\item \textbf{Human-Centered Protection Approach}
\begin{itemize}
    \item \textbf{Step-by-Step Verification}: Users verify outputs at each step before proceeding, ensuring early identification of errors or inconsistencies.This incremental process also reduces the task complexity thus improving the overall quality of the generated output.
    \item \textbf{Detailed Explanations with Outputs}: Generated artifacts include comprehensive explanations, allowing thorough review before integration.
    \item \textbf{Manual Integration into Workstreams}: Users manually integrate the generated outputs, providing an additional layer of scrutiny.
\end{itemize}

\item \textbf{Additional Techniques for Quality Assurance}
\begin{itemize}
  \item \textbf{Reverse Generation Verification:} This method involves feeding the generated output back into the LLM to verify its accuracy. For example, if an API is generated for a set of requirements, the API code is re-entered into the LLM with a prompt to generate the initial requirements. The two sets of requirements are then compared to ensure consistency and completeness as explained in [\cite{ponnusamy2025bridging}].

  \item \textbf{Utilization of Verification LLMs:} A secondary code specialized LLM is used to replicate the output generation process. The results from the primary and secondary models are compared to validate accuracy and quality.
\end{itemize}
\end{itemize}
Both reverse generation verification and the use of a verification LLM were applied selectively in the case study to assess the quality and completeness of outputs. The results showed good success in detecting and addressing potential errors. However, given the complexity of deploying and maintaining an additional LLM, \textbf{reverse generation verification} emerges as a more accessible and cost-effective option for enterprise deployments.\\

It is essential to emphasize that AI-generated code is also subject to the standard enterprise application development and delivery lifecycle. This lifecycle typically includes rigorous security checks across all stages, including development, application build, testing, deployment, and post-deployment. By combining these enterprise processes with the security measures outlined above during the initial code generation phase, the framework effectively addresses quality and security concerns associated with AI-generated code.
\section{CASE STUDY}
To showcase the capabilities of the proposed framework, the Spring PetClinic sample application was selected as the case study (\href{https://github.com/spring-projects/spring-petclinic/tree/main?tab=readme-ov-file}{Spring PetClinic GitHub Repository}). This application, built a few years ago using one of the most widely adopted Java frameworks, presents an ideal candidate due to its lack of formal documentation within the repository. The absence of comprehensive documentation mirrors real-world challenges often encountered during legacy application modernization.\\

By following the outlined framework step-by-step, the case study aimed to demonstrate the practicality, reliability, and effectiveness of the approach. The outputs generated during each phase and the corresponding insights and learnings are documented below.

\subsection{Requirements Generation for Legacy Applications}
\begin{itemize}
\item \textbf{Functional Requirements Generation for the Interaction Layer:}
\begin{itemize}
    \item Begin by loading the project structure from the local repository.
    \item Identify all files associated with the interaction layer.
    \item Generate functional requirements for each file within this layer.
    \item Carefully review the generated requirements to ensure accuracy and completeness and make necessary edits.
    \item Organize and save the reviewed requirements under the appropriate module or tag for clarity and future reference.
\end{itemize}

\vspace{-10pt}
\begin{figure}[H]
\centering
\includegraphics[width=0.75\textwidth]{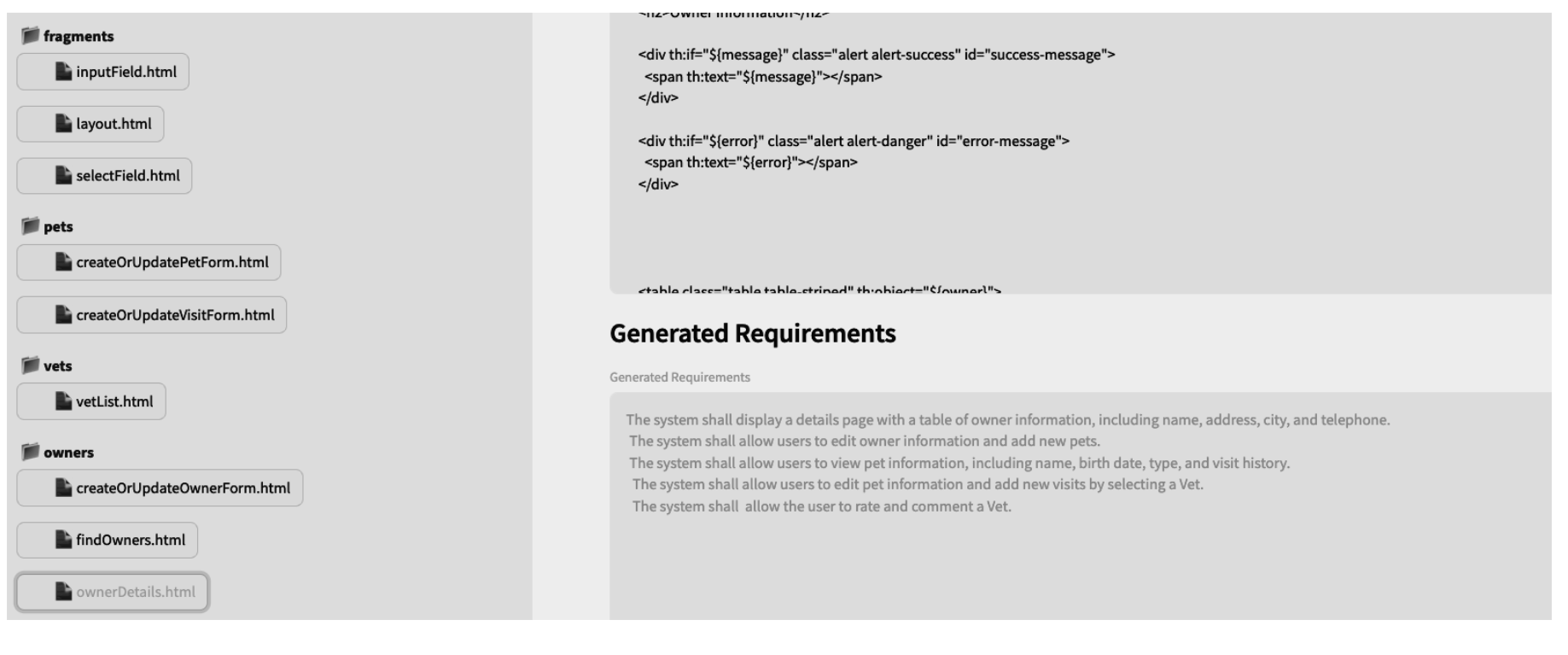}
\caption{Requirements generation output}
\label{fig:req_gen_output}
\end{figure}

\item \textbf{Consolidate Functional Requirements for Application/Module:}
\begin{itemize}
    \item Generate functional requirements for the desired module or the entire application.
    \item Review the generated output thoroughly and make necessary edits or additions to ensure completeness and accuracy.
     \item In this case study, new functional requirements were introduced to support a Veterinarian Rating feature, enhancing the application's capabilities. 
\end{itemize}
\vspace{-10pt}
\begin{figure}[H]
\centering
\includegraphics[width=0.75\textwidth]{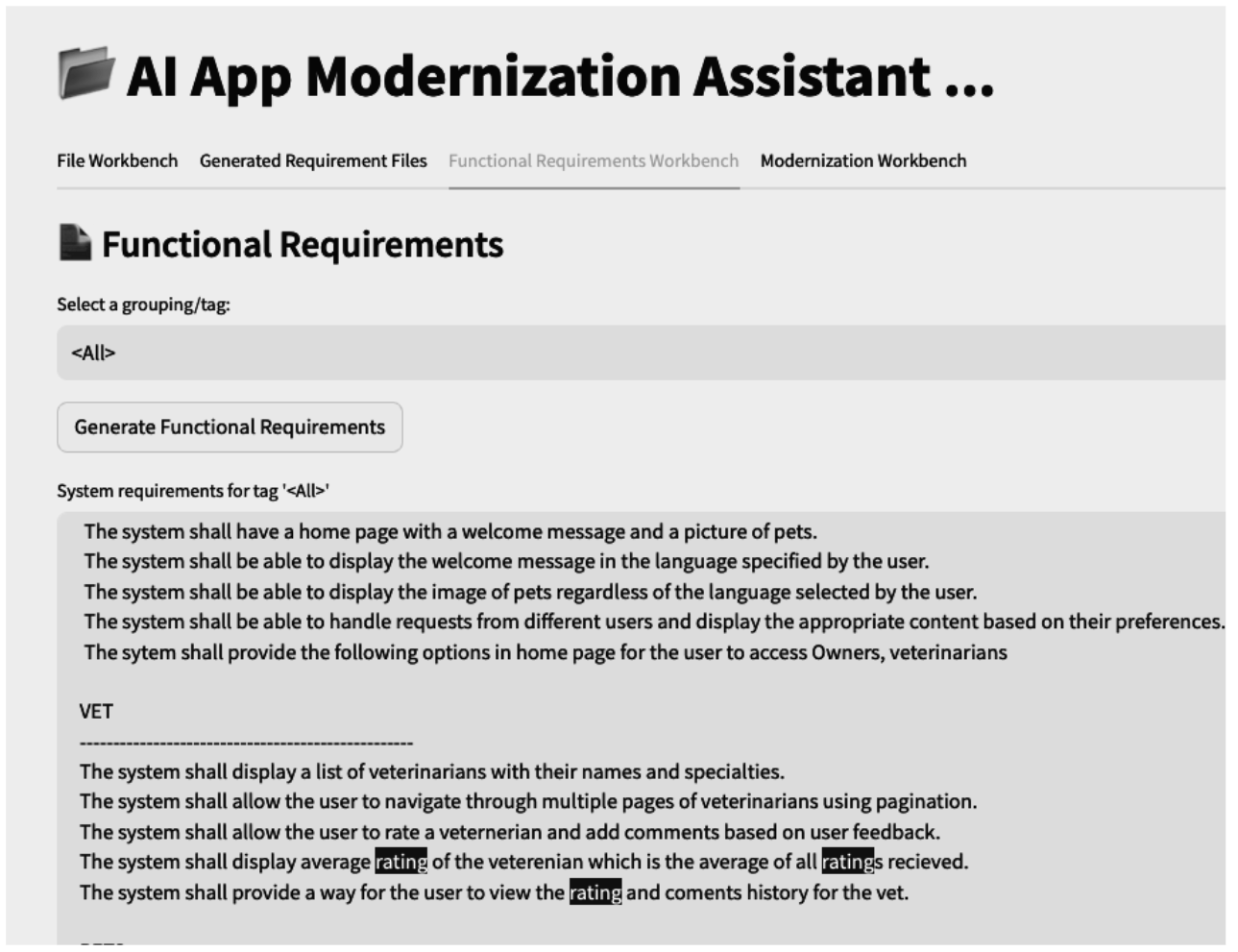}
\caption{Consolidated Application requirements with added ratings requirements.}
\label{fig:consolidate_reqs}
\end{figure}

With the requirements generated, reviewed, and refined with additional changes, the next step is to create artifacts for the modernized application by following the steps outlined in the framework. For the modernized application, Quarkus is selected as the application framework.  
\end{itemize}
\subsection{Development of a New Application Based on Refined Requirements}
\begin{itemize}
\item \textbf{Data Model and SQL Script Generation:}
\begin{itemize}
    \item Generate the Data Model and SQL script based on the application/module requirements.
    \item Review the generated output and accompanying explanations, making necessary adjustments to ensure accuracy and alignment with the requirements.
    \item If the output is incomplete or unsatisfactory, re-initiate the generation process to achieve improved results.
    \item Save the final output for future reference and execution as part of the development workflow.
\end{itemize}

\vspace{-10pt}
\begin{figure}[H]
\centering
\includegraphics[width=0.75\textwidth]{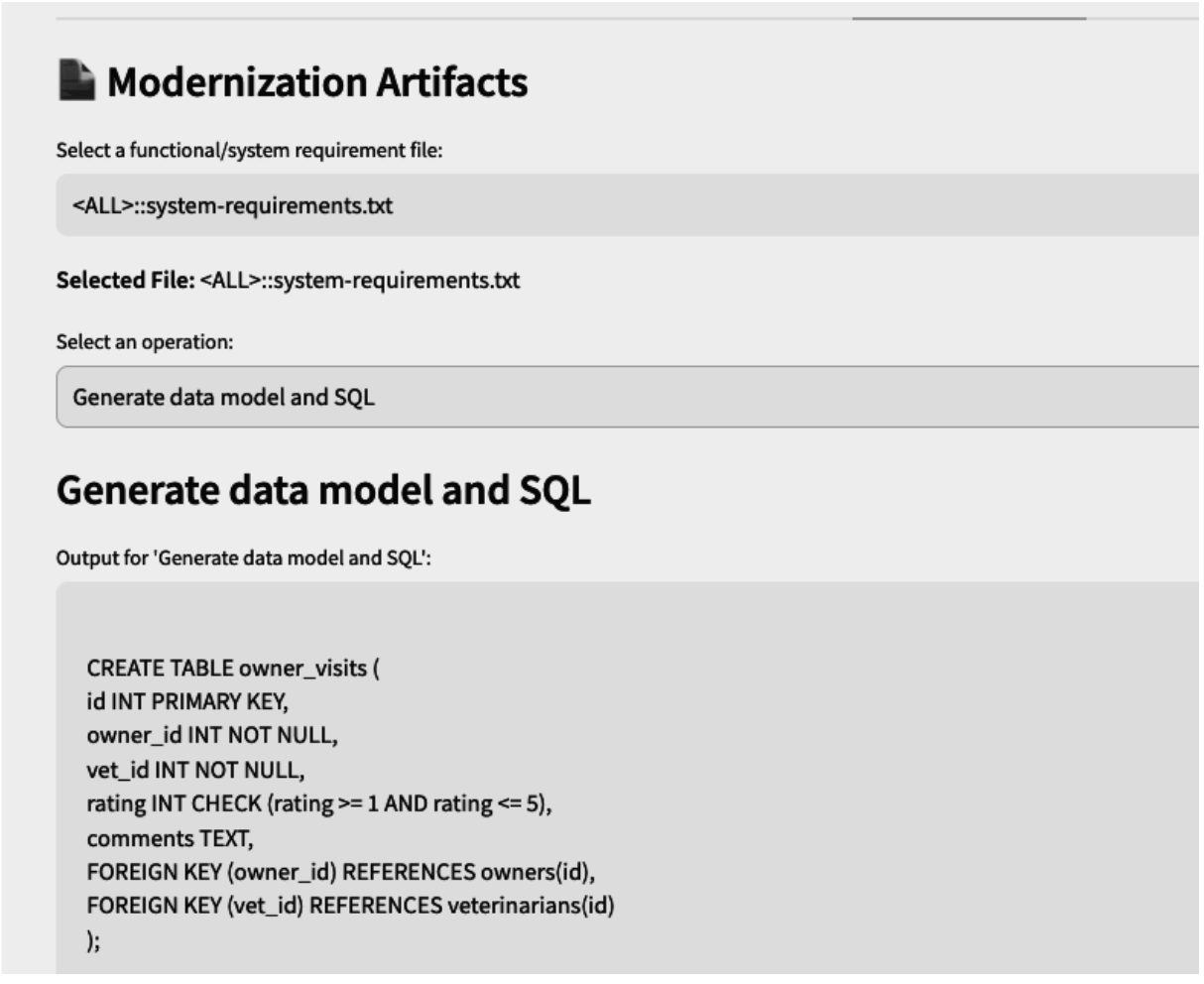}
\caption{Generated Data Model and SQL Script.}
\label{fig:generated_dm}
\end{figure}

\item \textbf{Generate Additional Artifacts:}
\begin{itemize}
    \item Generate ORM objects iteratively using the Data Model as the foundation.
    \item Utilize the ORM object code as input to create API code for the application.
    \item Based on the generated API code, produce comprehensive test cases to validate functionality.
    \item Leverage the functional requirements to create User Interface (UI) code, completing the application development workflow.
    \item Review the generated output and accompanying explanations, making necessary adjustments to ensure accuracy and alignment with the requirements.
    \item If the output is incomplete or unsatisfactory, re-initiate the generation process to achieve improved results.
    \item At every step, integrate code parts of the generated output into the modernization workstream as desired.
\end{itemize}
\end{itemize}

\section{OBSERVATIONS AND LIMITATIONS}

\begin{enumerate}[label=\alph*.]
	\item \textit{The reference implementation tool was specifically designed to meet the minimal requirements of the framework and is not intended for production environments. However, its modular structure makes it highly adaptable, allowing for easy extension and modification to suit specific needs.}

    \item \textit{During the exercise, occasional random behavior such as incomplete or misaligned responses was observed. Typically, simply resubmitting the request resolved this issue. This behavior was consistent across both open source and hosted models.}
    
    \item \textit{It was found that shorter input prompts generally yielded better output quality. This aligns with the framework's design, which generates artifacts incrementally to minimize prompt size and complexity. However, experiments with longer, random prompts—particularly when generating code and documentation—resulted in a noticeable decline in output quality.}
    
    \item \textit{The Reverse Generation Verification approach[\cite{ponnusamy2025bridging}] is a promising candidate for automation. However, due to the subtle variations in LLM outputs, creating a consistently reliable automation process proves challenging beyond relying on text similarity scores.}
    
    \item \textit{Random syntax errors were observed across all models during the exercise. The most effective way to address these was by either using IDE code assistant plugins or prompting another LLM to fix the issues. These two methods successfully mitigated syntax errors throughout the case study implementation.}
    
    \item \textit{The generated requirements documents do not align with the software engineering requirements templates proposed in \cite{wei2024requirements}. However, it is possible to convert the generated requirements into the desired templates, either manually or through automation.}
    
    \item \textit{Various quality and vulnerability analysis methods exist to enhance the quality and security of the generated code, as outlined in \cite{torka2024optimizing,taeb2024ai}. While implementing these methods can be costly and time-consuming with limited incremental value for this framework, they may offer significant benefits for enterprise-wide adoption, especially in highly regulated industries, justifying the higher costs.}
    
    \item \textit{While the reference implementation may have reduced the barrier to entry for experimentation and initial projects, scaling it to an enterprise-wide solution will require the deployment of an MLOps platform to enhance the availability, security, and management of the LLM, which serves as the foundation of this approach.}
    
    \item \textit{Although the case study focuses on functional requirements, tests conducted to generate non-functional requirements, such as performance requirements, demonstrated successful results.}
\end{enumerate}

\section{V. CONCLUSION}
This paper explores the advantages of utilizing Large Language Models (LLMs) to support application modernization efforts. By leveraging an open source code model, a structured framework, and a reference tool implementation, this paper outlines the steps involved in modernizing legacy applications. It emphasizes the transformative potential of LLMs when adopted in conjunction with a human review process, ensuring that automation and human oversight work collaboratively for optimal outcomes.\\

Additionally, the paper investigates various strategies for enhancing and securing the modernization approach, particularly for enterprise scale adoption. The discussion includes the importance of developing tools that improve both the user experience and the quality of outcomes across the enterprise, without necessitating advanced prompting skills from users.\\

The reference implementation tool, available on GitHub[\href{https://github.com/AhilanPonnusamy/App-Modernization-framework-reference-implementation}{GitHub Repository}], offers an open source solution that users can download, adapt, and extend to meet their specific needs. As LLMs continue to evolve and improve in their capabilities for code reasoning and generation, this framework, along with the associated tool, will have an even greater impact. Future iterations will be better equipped to address a broader range of enterprise use cases, making them invaluable for application modernization initiatives.

\bibliographystyle{IEEEtran}  
\bibliography{references}  

\end{document}